\documentclass[preprint,aps,amsmath,superscriptaddress,nofootinbib,tightenlines]{revtex4}
\usepackage{graphicx}
\usepackage{bm}
\usepackage{epsfig}
\usepackage[english]{babel}

\def\diag{\mbox{diag}}
\def\tr{\mbox{Tr}}

\begin{document}

\preprint{\vbox{\hbox{USTC-ICTS-08-12}}}

\title{From $N$ M2's to $N$ D2's}
\author{Yi Pang\footnote{Electronic address: yipang@itp.ac.cn}}
\affiliation{Interdisciplinary Center for Theoretical Study,\\
University of Science and Technology of China,\\
Hefei, Anhui 230026, China\\}
\affiliation{Institute of Theoretical Physics, Chinese Academy of Sciences,\\
P. O. Box 2735, Beijing 100080, China\\ \vspace{0.2cm}}
\author{Tower Wang\footnote{Electronic address: wangtao218@itp.ac.cn}}
\affiliation{Interdisciplinary Center for Theoretical Study,\\
University of Science and Technology of China,\\
Hefei, Anhui 230026, China\\}
\affiliation{Institute of Theoretical Physics, Chinese Academy of Sciences,\\
P. O. Box 2735, Beijing 100080, China\\ \vspace{0.2cm}}
\begin{abstract}
In this short note, we reduce the $\mathcal{N}=6$, $U(N)\times U(N)$
Chern-Simons gauge theories to $\mathcal{N}=8$, $U(N)$ Yang-Mills
gauge theories. This process corresponds to recovering the
world-volume theory of $N$ D2-branes from that of $N$ M2-branes in
an intermediate energy range. The supersymmetries are enhanced
because in this limit the branes localize far away from the orbifold
singularity. Our main scheme is exactly in accordance with Mukhi and
Papageorgakis's earlier work, although the Higgs mechanism becomes
trickier in the present case. We also speculate on applying the
scheme to a large class of new Bagger-Lambert models more generally.
\end{abstract}

\maketitle

\newpage



\section{Introduction}
In the past few months, the study of M2-branes was
revived\footnote{For a recent review of membrane, please refer to
\cite{Berman:2007bv}. Following the pioneer work
\cite{Bagger:2006sk}-\cite{Gustavsson:2008dy}, substantial
literature along this direction appears in the past half year. As
a partial list, see \cite{Bandres:2008vf}-\cite{Krishnan} and
references therein. We are sorry for missing a lot of relevant
work in our reference list.}, initiated by Bagger and Lambert
\cite{Bagger:2006sk,Bagger:2007jr,Bagger:2007vi} and Gustavsson
\cite{Gustavsson:2007vu,Gustavsson:2008dy}, in which the
world-volume theory of two M2-branes was discovered. However,
because the old Bagger-Lambert-Gustavsson (BLG) theory
\cite{Bagger:2006sk,Bagger:2007jr,Bagger:2007vi,Gustavsson:2007vu,Gustavsson:2008dy}
relies on a special form of 3-algebra, it is difficult to be
generalized to arbitrary number of M2-branes
\cite{Gustavsson:2008dy,Bandres:2008vf,Ho:2008bn,Papadopoulos:2008sk,Gauntlett:2008uf}.
Under certain conditions, the $\mathcal{A}_4$ algebra in the old
BLG theory is unique and there is a no-go theorem
\cite{Gauntlett:2008uf}. One can get around such a no-go theorem
by relaxing the conditions
\cite{Ho:2008bn,Gran:2008vi,Awata:1999dz}. Switching from
Euclidean 3-algebra to Lorentzian 3-algebra, three groups
\cite{Gomis:2008uv,Benvenuti:2008bt,Ho:2008ei} showed a possible
way for generalization. Around a month later, Aharony et al.
\cite{Aharony:2008ug} opened another possibility. Not restricted
to 3-algebra, they turn to consider Chern-Simons-matter theories
with a $U(N)\times U(N)$ or $SU(N)\times SU(N)$ gauge symmetry. In
fact, the old BLG theory can be reformulated into an $SU(2)\times
SU(2)$ theory
\cite{Bandres:2008vf,Mukhi:2008ux,VanRaamsdonk:2008ft,Berman:2008be,Distler:2008mk}.
What is more, it was shown long ago that the gauge group
$U(N)\times U(N)$ plays an important role in the infrared limit of
$N$ coincident D3-branes at a conical singularity
\cite{Klebanov:1998hh}. Therefore, an $SU(N)\times SU(N)$
Aharony-Bergman- Jafferis-Maldacena (ABJM) theory, as a
3-dimensional superconformal Chern-Simons-matter theory, is a very
attractive world-volume theory of $N$ coincident M2-branes. Given
that the old BLG theory is difficult for generalization, although
ABJM's $\mathcal{N}=6$ Chern-Simons-matter theories do not rely on
3-algebra, it is still appealing to recast them in terms of a
certain 3-algebra, in hopes that it may told us how to better
overcome the no-go theorem \cite{Gauntlett:2008uf}. This was
elegantly accomplished by Bagger and Lambert as reported in
\cite{Bagger:2008se} recently, which we will call the new BL
theory.

In this little exercise, we reduce the $\mathcal{N}=6$ $U(N)\times
U(N)$ Chern-Simons gauge theories to $U(N)$ Yang-Mills gauge
theories. We find the resulting theories have $\mathcal{N}=8$
supersymmetry, or 16 real supercharges. We will review ABJM theories
in section \ref{review}, and then in section \ref{Higgs} show the
details of Higgs mechanism for the present case. Finally, in section
\ref{conclusion}, we will comment on how to apply our scheme to the
new BL model generally and conclude.

\section{Review of $\mathcal{N}=6$ Chern-Simons theories}\label{review}
In this section, following the neat work \cite{Benna:2008zy}, we
review ABJM's $\mathcal{N}=6$, $U(N)\times U(N)$ Chern-Simons gauge
theories \cite{Aharony:2008ug}. We are interested in the special
case with an $SU(4)$ R-symmetry. In this case, the coefficients of
the Chern-Simons action and the superpotential in
\cite{Benna:2008zy} are related by $K=1/L$. Comparing with the old
BLG theory, one immediately reads $L=8\pi/k$ at level\footnote{An
example of Chern-Simons level quantization was shown in
\cite{Distler:2008mk}.} $k$. We mainly take the notations and
conventions of \cite{Benna:2008zy}, but the generators of $U(N)$
algebra are different in normalization,
\begin{eqnarray}\label{algebra}
\nonumber &&[T^{a},T^{b}]=if^{abc}T^{c},~~~~(a,b,c=1,2,...,N^2-1)\\
\nonumber &&\tr(T^{a}T^{b})=\frac{1}{2}\delta^{ab},~~~~\tr\{T^{a},T^{b}\}=\delta^{ab},~~~~\tr(\{T^{a},T^{b}\}T^{c})=\frac{1}{2}d^{abc},\\
&&T^{0}=\frac{1}{\sqrt{2N}}\diag(1,1,...,1)
\end{eqnarray}
in which we use the square bracket to denote commutators and the
brace bracket to denote anti-commutators. For convenience, we will
raise and lower group indices with a unit metric, so there is no
distinction among upper and lower indices, their position being
dictated by notational convenience. Nevertheless,  as usual,
repeated indices imply summation. The gauge potential $A_{\mu}$ and
$\tilde{A}_{\mu}$ in \cite{Benna:2008zy} will be replaced by
$A_{\mu}^{(L)}$ and $A_{\mu}^{(R)}$ in our notations, while the
covariant derivative $\mathcal{D}_{\mu}$ will be replaced by
$\tilde{\mathcal{D}}_{\mu}$, since we will use these notations for
other purposes.

After integrating out auxiliary fields and combining the $SU(2)$
fields into an $SU(4)$ representation,
\begin{eqnarray}
\nonumber Y^{1}&=&X^{1}+iX^{5},\\
\nonumber Y^{2}&=&X^{2}+iX^{6},\\
\nonumber Y^{3}&=&X^{3}+iX^{7},\\
Y^{4}&=&X^{4}+iX^{8},
\end{eqnarray}
one can write down the total action of ABJM's $\mathcal{N}=6$,
$U(N)\times U(N)$ Chern-Simons gauge theories as
\cite{Benna:2008zy,Bagger:2008se}
\begin{equation}\label{actionM2}
\mathcal{S}=\int d^3x\left[-\tr\left((\tilde{\mathcal{D}}^{\mu}Y_{A})^{\dag}\tilde{\mathcal{D}}_{\mu}Y^{A}\right)+i\tr\left(\psi^{A\dag}\gamma^{\mu}\tilde{\mathcal{D}}_{\mu}\psi_{A}\right)+\mathcal{L}_{\rm CS}-V^{\rm ferm}-V^{\rm bos}\right].
\end{equation}
In the above action, the covariant derivative is defined by
\begin{equation}
\tilde{\mathcal{D}_{\mu}}Y^{A}=\partial_{\mu}Y^{A}+iA_{\mu}^{(L)}Y^{A}-iY^{A}A_{\mu}^{(R)}.
\end{equation}
Note that the notation $\tilde{\mathcal{D}}_{\mu}$ with a
tilde here is different from $\mathcal{D}_{\mu}$ which will appear
later. The Chern-Simons term is
\begin{equation}
\mathcal{L}_{\rm CS}=\frac{2}{L}\epsilon^{\mu\nu\lambda}\tr\left(A_{\mu}^{(L)}\partial_{\nu}A_{\lambda}^{(L)}+\frac{2i}{3}A_{\mu}^{(L)}A_{\nu}^{(L)}A_{\lambda}^{(L)}-A_{\mu}^{(R)}\partial_{\nu}A_{\lambda}^{(R)}-\frac{2i}{3}A_{\mu}^{(R)}A_{\nu}^{(R)}A_{\lambda}^{(R)}\right).
\end{equation}
There is a potential like a Yukawa term, by which scalars and
fermions are coupled
\begin{eqnarray}
\nonumber V^{\rm ferm}&=&\frac{iL}{4}\tr\left(Y_{A}^{\dagger}Y^{A}\psi^{B\dagger}\psi_{B}-Y^{A}Y_{A}^{\dagger}\psi_{B}\psi^{B\dagger}+2Y^{A}Y_{B}^{\dagger}\psi_{A}\psi^{B\dagger}-2Y_{A}^{\dagger}Y^{B}\psi^{A\dagger}\psi_{B}\right.\\
&&\left.+\epsilon^{ABCD}Y_{A}^{\dagger}\psi_{B}Y_{C}^{\dagger}\psi_{D}-\epsilon_{ABCD}Y^{A}\psi^{B\dagger}Y^{C}\psi^{D\dagger}\right).
\end{eqnarray}
The sextic potential of scalars takes the form
\begin{eqnarray}
\nonumber V^{\rm bos}&=&-\frac{L^2}{48}\tr\left(Y^{A}Y_{A}^{\dagger}Y^{B}Y_{B}^{\dagger}Y^{C}Y_{C}^{\dagger}+Y_{A}^{\dagger}Y^{A}Y_{B}^{\dagger}Y^{B}Y_{C}^{\dagger}Y^{C}\right.\\
&&\left.+4Y^{A}Y_{B}^{\dagger}Y^{C}Y_{A}^{\dagger}Y^{B} Y_{C}^{\dagger}-6Y^{A}Y_{B}^{\dagger}Y^{B}Y_{A}^{\dagger}Y^{C}Y_{C}^{\dagger}\right).
\end{eqnarray}

\section{Reduction to Yang-Mills Theories}\label{Higgs}
In this section, we reduce the $U(N)\times U(N)$ Chern-Siomons
theories reviewed in the previous section into $U(N)$ Yang-Mills
theories. This process corresponds to recovering the world-volume
theory of $N$ D2-branes from that of $N$ M2-branes. Our main scheme
is exactly in accordance with \cite{Mukhi:2008ux}. Firstly recast
the gauge fields as\footnote{In the early days of preparing this
work, we found the choice (\ref{regauge}) was also made by the
reference \cite{Honma:2008jd}, which partly overlapped with our
plan. What is more, actually the scaling limit taken in
\cite{Honma:2008jd} is equivalent to the limit
$\tilde{v}\rightarrow\infty$ in our scheme. Anyhow we decided to
finish this little exercise to obtain more details.}
\begin{equation}\label{regauge}
A_{\mu}^{(L)a}=A_{\mu}^{a}+B_{\mu}^{a},~~~~A_{\mu}^{(R)a}=A_{\mu}^{a}-B_{\mu}^{a},
\end{equation}
and then integrate out the auxiliary field $B_{\mu}^{a}$ using its
equation of motion. At the same time, take the vacuum expectation
value to be
\begin{equation}\label{vev}
\langle Y^{A0}\rangle=i\tilde{v}\delta^{A}_{4}
\end{equation}
and other components vanished. $\tilde{v}$ denotes our choice of the
vacuum expectation value (VEV). In the following, $v$ denotes the choice in
\cite{Distler:2008mk}. We should stress here $\tilde{v}$ is real in
spite of the fact that $Y^{A0}$ is a complex variable. The
Yang-Mills coupling is defined by\footnote{It was argued in
\cite{Distler:2008mk} that $g_{YM}=v/\sqrt{k}$. At first glance, it
contradicts with the relation (\ref{gYM}). However, this
disagreement is understandable because the normalizations are
different. By the replacement $v\rightarrow4\pi
\tilde{v}/\sqrt{2kN}$, one can change from their normalization to
ours and recover (\ref{gYM}). In the last step of (\ref{gYM}), we
have used the quantization condition $L=8\pi/k$.}
\begin{equation}\label{gYM}
g_{YM}=\frac{L\tilde{v}}{2\sqrt{2N}}=\frac{4\pi \tilde{v}}{k\sqrt{2N}}
\end{equation}
as explained in \cite{Distler:2008mk}. As did in
\cite{Mukhi:2008ux,Distler:2008mk}, we have to rescale some of the
components of $Y$ and $\psi$ as
$(X,\psi)\rightarrow(X/g_{YM},\psi/g_{YM})$ and keep only leading
order terms with respect to $\tilde{v}^{-1}$. In other words, we
have to take the limit $k\rightarrow\infty$,
$\tilde{v}\rightarrow\infty$ with $g_{YM}$ fixed. One possible
interpretation of this limit is illustrated by Figure 1.
\begin{figure}[ht]
\begin{center}
\includegraphics[width=0.6\textwidth]{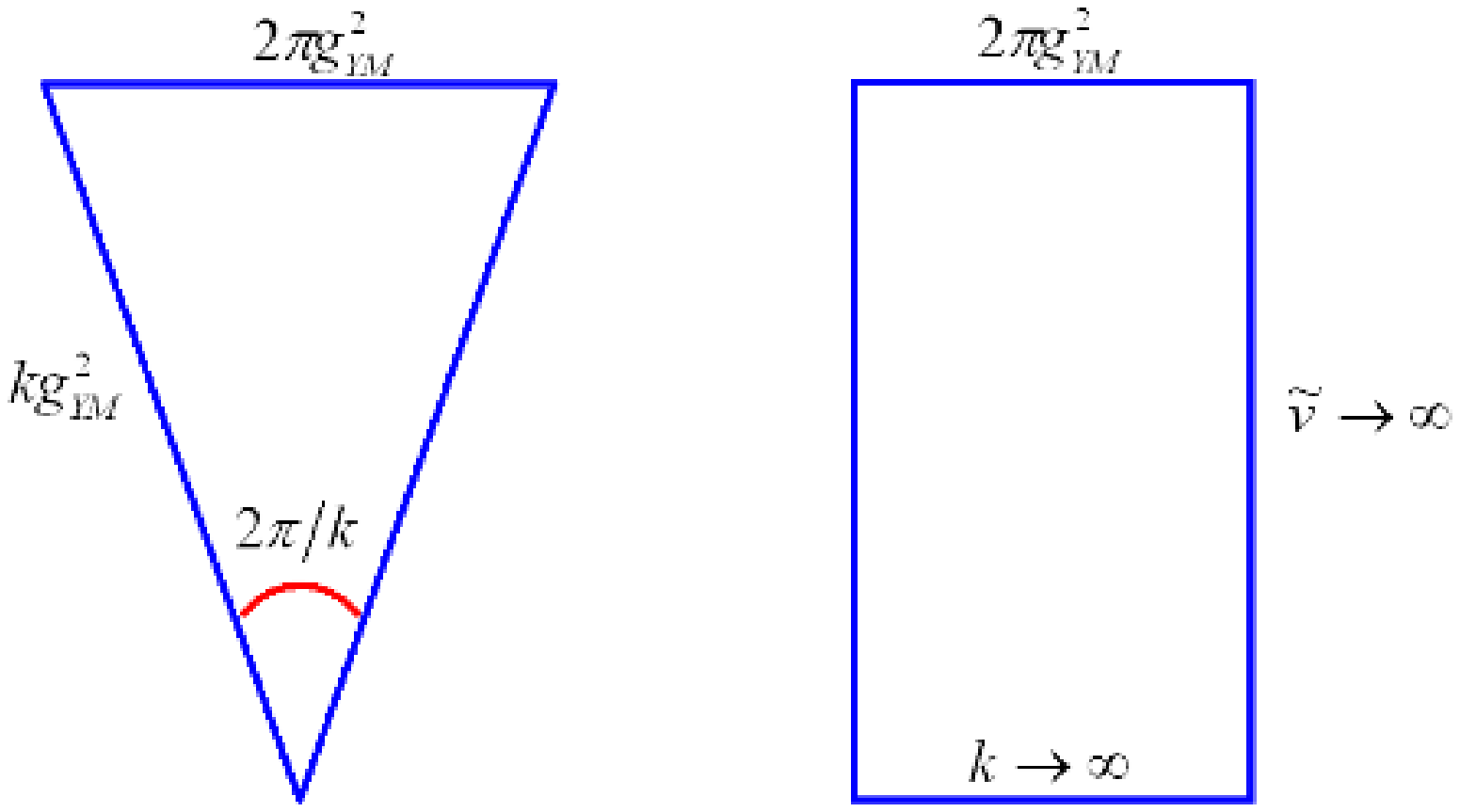}
\end{center}
\caption{The left figure signifies the $C^{4}/Z_{k}$. In the rest of
our paper, we will see that $kg_{YM}^{2}$ is a typical energy scale,
below which our reduction is meaningful. According to the definition
of $g_{YM}$, $kg_{YM}^{2}\sim \tilde{v}^{2}/(kN)\sim v^{2}$. Since
in reference \cite{Distler:2008mk} $v^{2}$ is also related to
physical mass scale, we see again that these two descriptions are
equivalent. So taking the limit $k\rightarrow\infty$ makes the cone
become a fine cylinder. It is the usual picture that by
compactifying the $x^{11}$, M-theory is reduced to the type IIA
string theory.}
\end{figure}

Many of the techniques employed here can be found in
\cite{Mukhi:2008ux,Honma:2008jd}. In reference \cite{Honma:2008jd},
a precise relation between ABJM model and Lorentzian BLG model was
studied. Since the Lorentzian BLG model is intimately related to D2
branes, our will  partly overlap with the analysis there. In their
analysis, they promoted the coupling constant to a spacetime
dependent vector and recovered the $SO(8)$ invariance. Since our aim
is to get the world-volume theory of D2-branes, we will take
$\tilde{v}$ to be spacetime independent and get the $SO(7)$
symmetry.

First of all, we decompose fields as
\begin{equation}
Y^{A}=Y^{A0}T^{0}+iY^{Aa}T^{a},~~~~\psi_{A}=\psi_{A}^{0}T^{0}+i\psi_{A}^{a}T^{a},~~~~A_{\mu}=A_{\mu}^{a}T^{a}.
\end{equation}
Here $Y^{A0}$ and $Y^{Aa}$ are complex variables, $\psi_{A}^{0}$ and
$\psi_{A}^{a}$ are complex with two components, while $A_{\mu}^{a}$
is a real variable. Let us also introduce
\begin{eqnarray}
\nonumber \mathcal{D}_{\mu}Y^{A0}&=&\partial_{\mu}Y^{A0},\\
\nonumber \mathcal{D}_{\mu}Y^{Aa}&=&\partial_{\mu}Y^{Aa}-f^{abc}A_{\mu}^{b}Y^{Ac},\\
\nonumber (\mathcal{D}_{\mu}Y_{A}^{a})^{\dag}&=&(\mathcal{D}_{\mu}Y_{A}^{a})^{*}=\partial_{\mu}\bar{Y}_{A}^{a}-f^{abc}A_{\mu}^{b}\bar{Y}_{A}^{c},\\
F_{\nu\lambda}^{a}&=&\partial_{\nu}A^{a}_{\lambda}-\partial_{\lambda}A^{a}_{\nu}-f^{abc}A_{\nu}^{b}A_{\lambda}^{c}.
\end{eqnarray}

Making use of
\begin{equation}
\tilde{\mathcal{D}}_{\mu}Y^{A}=\partial_{\mu}Y^{A}+i[A_{\mu},Y^{A}]+i\{B_{\mu},Y^{A}\}=\mathcal{D}_{\mu}Y^{A}+i\{B_{\mu},Y^{A}\},
\end{equation}
and conventions (\ref{algebra}), one can rewrite the kinetic term of
scalars in the form
\begin{eqnarray}\label{kinbos2}
\nonumber &&-\tr\left((\tilde{\mathcal{D}}^{\mu}Y_{A})^{\dag}\tilde{\mathcal{D}}_{\mu}Y^{A}\right)\\
\nonumber &=&-\frac{1}{2}\partial^{\mu}\bar{Y}_{A}^{0}\partial_{\mu}Y^{A0}-\frac{1}{2}(\mathcal{D}^{\mu}Y^{Aa})^{\dag}\mathcal{D}_{\mu}Y^{Aa}-\frac{1}{N}B_{\mu}^{a}B^{\mu a}\bar{Y}_{A}^{0}Y^{A0}\\
\nonumber &&+\frac{i}{\sqrt{2N}}d^{abc}B_{\mu}^{a}B^{\mu b}(Y^{A0}\bar{Y}_{A}^{c}-\bar{Y}_{A}^{0}Y^{Ac})-B_{\mu}^{a}B^{\mu b}\bar{Y}_{A}^{c}Y^{Ad}\tr\left(\{T^{a},T^{c}\}\{T^{b},T^{d}\}\right)\\
\nonumber &&-\frac{1}{\sqrt{2N}}B^{\mu a}\left[(\mathcal{D}_{\mu}Y^{Aa})\bar{Y}_{A}^{0}+(\mathcal{D}_{\mu}Y_{A}^{a})^{\dag}Y^{A0}\right]\\
\nonumber &&+\frac{1}{\sqrt{2N}}B^{\mu a}\left[(\partial_{\mu}Y^{A0})\bar{Y}_{A}^{a}+(\partial_{\mu}\bar{Y}_{A}^{0})Y^{Aa}\right]\\
&&+\frac{i}{2}d^{abc}B^{\mu a}\left[(\mathcal{D}_{\mu}Y^{Ab})\bar{Y}_{A}^{c}-(\mathcal{D}_{\mu}Y_{A}^{b})^{\dag}Y^{Ac}\right],
\end{eqnarray}
while the kinetic term of fermions takes the form
\begin{eqnarray}\label{kinferm2}
\nonumber &&i\tr\left(\psi^{A\dag}\gamma^{\mu}\tilde{\mathcal{D}}_{\mu}\psi_{A}\right)\\
\nonumber &=&\frac{i}{2}(\psi^{A0\dag}\gamma^{\mu}\partial_{\mu}\psi_{A}^{0}+\psi^{Aa\dag}\gamma^{\mu}\mathcal{D}_{\mu}\psi_{A}^{a})\\
&&+\frac{i}{2N}B_{\mu}^{a}(\psi^{Aa\dag}\gamma^{\mu}\psi_{A}^{0}-\psi^{A0\dag}\gamma^{\mu}\psi_{A}^{a})-\frac{1}{2}d^{abc}B_{\mu}^{a}\psi^{Ab\dag}\gamma^{\mu}\psi_{A}^{c}.
\end{eqnarray}

As for the Chern-Simons term, we obtain
\begin{eqnarray}\label{chesim2}
\mathcal{L}_{\rm CS}&=&\frac{2}{L}\epsilon^{\mu\nu\lambda}(B_{\mu}^{a}F_{\nu\lambda}^{a}-\frac{1}{3}f^{abc}B_{\mu}^{a}B_{\nu}^{b}B_{\lambda}^{c}).
\end{eqnarray}
Although the gauge fields $A_{\mu}^{(L,R)}$ are purely topological,
after the recombination (\ref{regauge}), one of the new fields
$B_{\mu}^{a}$ becomes an auxiliary field without derivatives, just
as happened in \cite{Mukhi:2008ux}. We will eliminate this auxiliary
field using its equation of motion, and obtain the kinetic term of
the other gauge field $A_{\mu}^{a}$. Namely, $A_{\mu}^{a}$ will get
dynamical if one integrates out $B_{\mu}^{a}$.

The potential term mixing scalars and fermions becomes
\begin{eqnarray}\label{Vferm2}
\nonumber V^{\rm ferm}&=&\frac{iL}{4}\tr\biggl[\frac{1}{2\sqrt{2N}}f^{abc}(Y^{A0}\bar{Y}_{A}^{a}-\bar{Y}_{A}^{0}Y^{Aa})\bar{\psi}^{Bb}\psi_{B}^{c}\\
\nonumber &&+\frac{1}{2\sqrt{2N}}f^{abc}\bar{Y}_{A}^{a}Y^{Ab}(\bar{\psi}^{Bc}\psi_{B}^{0}-\psi^{Bc}\bar{\psi}^{B0})\\
\nonumber &&+\frac{1}{\sqrt{2N}}f^{abc}(Y^{A0}\bar{Y}_{B}^{a}\psi_{A}^{b}\bar{\psi}^{Bc}+\bar{Y}_{A}^{0}Y^{Ba}\bar{\psi}^{Ab}\psi_{B}^{c})\\
\nonumber &&-\frac{1}{\sqrt{2N}}f^{abc}(Y^{Aa}\bar{Y}_{B}^{b}\psi_{A}^{c}\bar{\psi}^{B0}+\bar{Y}_{A}^{a}Y^{Bb}\bar{\psi}^{Ac}\psi_{B}^{0})\\
\nonumber &&+(\bar{Y}_{A}^{a}Y^{Ab}\bar{\psi}^{Bc}\psi_{B}^{d}+2Y^{Aa}\bar{Y}_{B}^{b}\psi_{A}^{c}\bar{\psi}^{Bd})\tr\left(T^{a}T^{b}T^{c}T^{d}-T^{b}T^{a}T^{d}T^{c}\right)\\
\nonumber &&-\frac{1}{2\sqrt{2N}}f^{abc}(\epsilon^{ABCD}\bar{Y}_{A}^{0}\psi_{B}^{a}\bar{Y}_{C}^{b}\psi_{D}^{c}+\epsilon_{ABCD}Y^{A0}\bar{\psi}^{Ba}Y^{Cb}\bar{\psi}^{Dc})\\
\nonumber &&+\frac{1}{2\sqrt{2N}}f^{abc}(\epsilon^{ABCD}\bar{Y}_{A}^{a}\psi_{B}^{b}\bar{Y}_{C}^{c}\psi_{D}^{0}+\epsilon_{ABCD}Y^{Aa}\bar{\psi}^{Bb}Y^{Cc}\bar{\psi}^{D0})\\
&&+(\epsilon^{ABCD}\bar{Y}_{A}^{a}\psi_{B}^{b}\bar{Y}_{C}^{c}\psi_{D}^{d}-\epsilon_{ABCD}Y^{Aa}\bar{\psi}^{Bb}Y^{Cc}\bar{\psi}^{Dd})\tr\left(T^{a}T^{b}T^{c}T^{d}\right)\biggr].
\end{eqnarray}
The bosonic potential can be written as
\begin{eqnarray}\label{Vbos2}
\nonumber V^{\rm bos}&=&-\frac{L^2}{48}\biggl[\frac{3}{2N}Y^{A0}\bar{Y}_{A}^{0}Y^{Ba}\bar{Y}_{B}^{b}Y^{Cc}\bar{Y}_{C}^{d}\tr\Bigl([T^{a},T^{d}][T^{b},T^{c}]+[T^{a},T^{c}][T^{b},T^{d}]\Bigr)\\
\nonumber &&-\frac{3}{2N}(Y^{A0}Y^{B0}\bar{Y}_{A}^{a}\bar{Y}_{B}^{b}+\bar{Y}_{A}^{0}\bar{Y}_{B}^{0}Y^{Aa}Y^{Bb})Y^{Cc}\bar{Y}_{C}^{d}\tr\Bigl([T^{a},T^{c}][T^{b},T^{d}]\Bigr)\\
\nonumber &&-\frac{3}{2N}Y^{A0}\bar{Y}_{B}^{0}\bar{Y}_{A}^{a}Y^{Bb}Y^{Cc}\bar{Y}_{C}^{d}\tr\Bigl([T^{a},T^{d}][T^{b},T^{c}]+[T^{a},T^{c}][T^{b},T^{d}]\Bigr)\\
\nonumber &&+\frac{3i}{\sqrt{2N}}(\bar{Y}_{A}^{0}Y^{Aa}Y^{Bb}\bar{Y}_{B}^{c}Y^{Cd}\bar{Y}_{C}^{e}-Y^{A0}\bar{Y}_{A}^{a}\bar{Y}_{B}^{b}Y^{Bc}\bar{Y}_{C}^{d}Y^{Ce})\\
\nonumber &&\times\tr\Bigl(2(T^{b}T^{e}T^{a}-T^{a}T^{e}T^{b})[T^{c},T^{d}]+(T^{d}T^{e}T^{a}-T^{a}T^{e}T^{d})[T^{b},T^{c}]\Bigr)\\
\nonumber &&+Y^{Aa}\bar{Y}_{A}^{b}Y^{Bc}\bar{Y}_{B}^{d}Y^{Ce}\bar{Y}_{C}^{f}\tr\Bigl(T^{a}T^{b}T^{c}T^{d}T^{e}T^{f}+T^{b}T^{a}T^{d}T^{c}T^{f}T^{e}\\
&&+4T^{a}T^{d}T^{e}T^{b}T^{c}T^{f}-6T^{a}T^{d}T^{c}T^{b}T^{e}T^{f}\Bigr)\biggr].
\end{eqnarray}

Choosing the vacuum expectation value given in (\ref{vev}), we can
expand the scalars and fermions near the VEV as
\begin{eqnarray}
\nonumber Y^{A}&=&x^{A}_{0}T^{0}+i(\tilde{v}\delta^{A}_{4}+x^{A+4}_{0})T^{0}+ix^{Aa}T^{a}-x^{(A+4)a}T^{a},\\
\psi^{A}&=&\Psi^{A}_{0}T^{0}+i\Psi^{A+4}_{0}T^{0}+i\Psi^{Aa}T^{a}-\Psi^{(A+4)a}T^{a}.
\end{eqnarray}

In the limit $\tilde{v}\rightarrow\infty$, the leading order terms in
(\ref{actionM2}) can be obtained using the above results. In
particular, the leading order kinetic terms in (\ref{kinbos2}) and
(\ref{kinferm2}) give
\begin{eqnarray}
\nonumber \mathcal {L}_{\rm kinetic}&=&-\sum_{I=1}^{8}\frac{1}{2}\left(\partial^{\mu}x^{I}_{0}\partial_{\mu}x^{I}_{0}+\mathcal{D}^{\mu}x^{Ia}\mathcal{D}_{\mu}x^{Ia}\right)-\frac{\tilde{v}^{2}}{N}B_{\mu}^{a}B^{\mu a}-\frac{2\tilde{v}}{\sqrt{2N}}B^{\mu a}\mathcal{D}_{\mu}x^{8a}\\
\nonumber &&+\sum_{A=1}^{4}\frac{i}{2}(\Psi^{A\dag}_{0}-i\Psi^{(A+4)\dag}_{0})\gamma^{\mu}\partial_{\mu}(\Psi^{A}_{0}+i\Psi^{A+4}_{0})\\
&&+\sum_{A=1}^{4}\frac{i}{2}(\Psi^{Aa\dag}-i\Psi^{(A+4)a\dag})\gamma^{\mu}\mathcal{D}_{\mu}(\Psi^{Aa}+i\Psi^{(A+4)a}).
\end{eqnarray}
As we will show later, the cubic term in (\ref{chesim2}) is
negligible in the limit $\tilde{v}\rightarrow\infty$, so only the first
term survives
\begin{equation}
\mathcal{L}_{\rm CS}=\frac{2}{L}\epsilon^{\mu\nu\lambda}B_{\mu}^{a}F_{\nu\lambda}^{a}.
\end{equation}

The gauge potential $B_{\mu}^{a}$ appears only in $\mathcal {L}_{\rm
kinetic}$ and $\mathcal{L}_{\rm CS}$ without derivatives. We can
eliminate it using the equation of motion
\begin{equation}
B^{\mu a}=\frac{N}{L\tilde{v}^2}\epsilon^{\mu\nu\lambda}F_{\nu\lambda}^{a}-\frac{\sqrt{2N}}{2\tilde{v}}\mathcal{D}^{\mu}x^{8a}.
\end{equation}
When deriving this equation we have neglected the quadratic term in
$B^{\mu a}$ coming from the cubic self-interaction as well as terms
coming from higher interaction with scalars. Because these terms
will lead to higher order contributions which are suppressed in the
limit $\tilde{v}\rightarrow\infty,L\rightarrow0$.

Inserting the above equation into the lagrangian, up to a total
derivative term, the kinetic and Chern-Simons terms turn out to be
\begin{eqnarray}\label{kin3}
\nonumber &&\mathcal {L}_{\rm kinetic}+\mathcal{L}_{\rm CS}\\
\nonumber &=&-\frac{2N}{L^2\tilde{v}^2}F^{\nu\lambda a}F_{\nu\lambda}^{a}-\sum_{j=1}^{7}\frac{1}{2}\left(\partial^{\mu}x^{j}_{0}\partial_{\mu}x^{j}_{0}+\mathcal{D}^{\mu}x^{ja}\mathcal{D}_{\mu}x^{ja}\right)-\frac{1}{2}\partial^{\mu}x^{8}_{0}\partial_{\mu}x^{8}_{0}\\
\nonumber &&+\sum_{A=1}^{4}\frac{i}{2}(\Psi^{A\dag}_{0}-i\Psi^{(A+4)\dag}_{0})\gamma^{\mu}\partial_{\mu}(\Psi^{A}_{0}+i\Psi^{A+4}_{0})\\
&&+\sum_{A=1}^{4}\frac{i}{2}(\Psi^{Aa\dag}-i\Psi^{(A+4)a\dag})\gamma^{\mu}\mathcal{D}_{\mu}(\Psi^{Aa}+i\Psi^{(A+4)a}),
\end{eqnarray}
We find the kenetic terms of $x^{8a}$ exactly cancel out; hence
they become non-dynamical. But a question may arise: where are the
degrees of freedom newly turned on to compensate the disappeared
$x^{8a}$. The doubt can be resolved by observing that the gauge
field $A_{\mu}^{a}$ becomes dynamical. Since in three dimensions,
each massless vector field $A_{\mu}^{a}$ has one dynamical degree
of freedom as the scalar $x^{8a}$, the total degrees of freedom
are unchanged. From the Yang-Mills term, by requiring its
coefficient to be $-1/(4g_{YM}^2)$, one can quickly read the
Yang-Mills coupling constant as in (\ref{gYM}). It is clear that
$B^{\mu a}\propto L/g_{YM}^2$ right now. For a finite $g_{YM}$, in
the limit $L\rightarrow0$, the first term in (\ref{chesim2}) is
finite while the second term vanishes, as we have assumed. Since
this limit is to keep the leading term in an expansion in powers
of energy divided by $\tilde{v}^{2}/k$, this also tells us that
the resulting $U(N)$ gauge theory is valid only at a energy below
$g_{YM}^2/L$ i.e. $kg_{YM}^2$. On the other hand, since our
derivation is to depict the weakly coupled limit of N D-branes,
the theory is applicable only above the energy scale $g_{YM}^2N$.

In the limit $\tilde{v}\rightarrow\infty,L\rightarrow0$ while keeping
$L\tilde{v}$ finite, the potential terms for scalars and fermions reduce
to
\begin{eqnarray}\label{Vferm3}
\nonumber V^{\rm ferm}&=&-\frac{L\tilde{v}}{4\sqrt{2N}}f^{abc}\tr\biggl[\sum_{A=1}^{3}2ix^{4a}\Psi^{Ab}\Psi^{(A+4)c}-2ix^{4a}\Psi^{4b}\Psi^{8c}\\
\nonumber &&+\sum_{A=1}^{3}2ix^{Aa}(\Psi^{8b}\Psi^{Ac}-\Psi^{4b}\Psi^{(A+4)c})-\sum_{A=1}^{3}2ix^{(A+4)a}(\Psi^{4b}\Psi^{Ac}+\Psi^{8b}\Psi^{(A+4)c})\\
\nonumber &&-\sum_{A,B,C=1}^{3}\frac{1}{2}\epsilon^{ABC4}(\Psi^{Aa}+i\Psi^{(A+4)a})(x^{Bb}-ix^{(B+4)b})(\Psi^{Cc}+i\Psi^{(C+4)c})\\
\nonumber &&+\sum_{A,B,C=1}^{3}\frac{1}{2}\epsilon_{ABC4}(\Psi^{Aa}-i\Psi^{(A+4)a})(x^{Bb}+ix^{(B+4)b})(\Psi^{Cc}-i\Psi^{(C+4)c})\biggr].\\
\end{eqnarray}
\begin{equation}\label{Vbos3}
V^{\rm
bos}=-\sum_{i,j=1}^{7}\frac{L^2\tilde{v}^2}{16N}x^{ia}x^{ib}x^{jc}x^{jd}\tr\left([T^{a},T^{c}][T^{b},T^{d}]\right).
\end{equation}
In (\ref{Vferm2}) and (\ref{Vferm3}), we keep the notation $\tr$ to
remind the inner product between two-component spinors. As one
expected, the goldstones $x^{8a}$ disappear in
(\ref{kin3}-\ref{Vbos3}) and hence in the total action.

If we define the minimal spinor of $SO(2,1)\times SO(7)$ as,
\begin{equation}
\Psi=\begin{pmatrix}
\Psi^{1a}, & \Psi^{2a}, & \Psi^{3a}, & \Psi^{4a}, & \Psi^{5a}, & \Psi^{6a}, & \Psi^{7a}, & \Psi^{8a} \\
\end{pmatrix}^{T},
\end{equation}
where each $\Psi^{i}$ is also a two-component Majorana spinor. Then
the fermionic potential is actually the standard Yukawa coupling:
\begin{equation}
 V^{\rm ferm}=-\frac{L\tilde{v}}{4\sqrt{2N}}f^{abc}x^{ia}\bar{\Psi}^{b}\Gamma^{i}\otimes\mathbf{1}_{2\times2}\Psi^{c}.
\end{equation}
At this moment, it is easy to see that the fermionic potential
term also has an $SO(7)$ R-symmetry.

The last step is to rescale the scalars and spinor as
$(x^{j},\Psi)\rightarrow(x^{j}/g_{YM},\Psi/g_{YM})$. Gathering the
above results together, in the limit $\tilde{v}\rightarrow\infty$ with a
fixed $g_{YM}$, we have
\begin{equation}\label{actionD2}
\mathcal{S}=\int d^3x\left(\mathcal{L}_{\rm
decoupled}+\frac{1}{g_{YM}^2}\mathcal{L}_{\rm coupled}\right),
\end{equation}
where
\begin{equation}
\mathcal{L}_{\rm decoupled}=-\sum_{j=1}^{7}\frac{1}{2}\partial^{\mu}x^{j}_{0}\partial_{\mu}x^{j}_{0}+\frac{i}{2}\Psi_{0}\Gamma^{0}\otimes\gamma^{\mu}\partial_{\mu}\Psi_{0}-\frac{1}{2}\partial^{\mu}x^{8}_{0}\partial_{\mu}x^{8}_{0},
\end{equation}
\begin{eqnarray}
\nonumber \mathcal{L}_{\rm coupled}&=&-\frac{1}{4}F^{\nu\lambda a}F_{\nu\lambda}^{a}-\frac{1}{2}\mathcal{D}^{\mu}x^{ja}\mathcal{D}_{\mu}x^{ja}+\frac{i}{2}\Psi^{a}\Gamma^{0}\otimes\gamma^{\mu}\mathcal{D}_{\mu}\Psi_{a}\\
&&-\frac{1}{2}\tr\left([x^{i},x^{j}][x^{i},x^{j}]\right)-\frac{1}{2}f^{abc}x^{ja}\bar{\Psi}^{b}\Gamma^{j}\otimes\mathbf{1}\Psi^{c}.
\end{eqnarray}
As one should have expected, at last we get a decoupled $U(1)$
sector and a coupled $SU(N)$ sector, which is nothing else but an
$\mathcal{N}=8$ super Yang-Mills theory on the world-volume of $N$
coincident D2-branes. The action (\ref{actionD2}) is valid in the
energy range between $g_{YM}^2N$ and $kg_{YM}^2$, and we have
assumed $N\ll k$ in the above procedure.

\section{Discussion of New BL model and Conclusion}\label{conclusion}
In new BL models \cite{Bagger:2008se}, in general the structure
constant of 3-algebra does not have three totally anti-symmetric
indices, so one cannot apply Mukhi and Papageorgakis's method
\cite{Mukhi:2008ux} in a naive way. This difficulty has been
discussed in a newly appeared paper \cite{Cherkis:2008qr}, where
Cherkis and Saemann also proposed a class of models similar to (but
a little different from) the new BL models.

One viable extrapolation of Mukhi and Papageorgakis's method
\cite{Mukhi:2008ux} is to decompose the 3-algebra into a couple of
2-algebras, say $\mathcal{G}_1\oplus\mathcal{G}_2$, just as did
in \cite{Bagger:2008se}:
\begin{equation}
f^{abcd}=\sum_{\lambda}\omega_{\lambda}\sum_{\alpha}(t^{\alpha}_{\lambda})^{ad}(t^{\alpha}_{\lambda})^{bc}.
\end{equation}
When $\mathcal{G}_1=\mathcal{G}_2=u(N)$, it exactly recovers the
ABJM theories, and the Higgs mechanism we discussed in the previous
section exactly applies.

More generally, if $\mathcal{G}_1=\mathcal{G}_2$ is another Lie
algebra instead of $u(N)$, the mechanism is similar. Some key points
are:
\begin{enumerate}
\item taking the VEV as (\ref{vev}), which has also been mentioned in
\cite{Aharony:2008ug};
\item rearrange the gauge field as (\ref{regauge});
\item decompose four complex scalar (spinor) fields into eight real
fields.
\end{enumerate}
The procedure shown in the previous section still works if only
$\mathcal{G}_1=\mathcal{G}_2$, although that field theory would have
less supersymmetries. It is easy to see that in general the
goldstones becomes non-dynamical, at least they will disappear in
the kinetic term. While the $G_1\times G_1$ gauge symmetry is broken
down to $G_1$. The degrees of freedom of the goldstone bosons are
transferred to the dynamical $G_1$ gauge fields.

However, if $\mathcal{G}_1\neq\mathcal{G}_2$, the above scheme is
helpless and things would be much trickier.

As a conclusion, we can see that Mukhi and Papageorgakis's method,
after a little adjustment, is still powerful in deriving the
world-volume action of $N$ D2-branes from that of $N$ M2-branes when
$N$ is arbitrary.

\emph{Note added}: When this work was finished, a similar paper
\cite{Li:2008ya} appeared with the emphasis on the pure Yang-Mils
terms. The interested readers are encouraged to compare our results
with that paper, especially the Yang-Mills coupling constant.

\acknowledgments{We are grateful to Miao Li for many helpful
comments. Also we would like to thank Wei He, Yushu Song and Gang
Yang for discussions. This work was supported by grants of CNSF and
grants of USTC.}

\end{document}